\documentclass[aps, prl, reprint, amsfont, amssymb, amsmath]{revtex4-1}

\usepackage{amsmath}
\usepackage{graphicx}
\usepackage{mathtools}
\usepackage{xcolor}
\usepackage{todonotes}
\usepackage{hyperref}
\begin{document}

\bibliographystyle{apsrev4-1}

\renewcommand{\refname}{References}

\title{Ferromagnetic CrBr$_3$-Induced Graphene Spintronics}
\author{Sushant Kumar Behera}
\affiliation{Materials Science Division, Lawrence Berkeley National Laboratory, Berkeley 94720 CA, USA}
\affiliation{Materials Engineering, Indian Institute of Science Bengaluru 560012, India.}

\author{Praveen C Ramamurthy}
\affiliation{Materials Engineering, Indian Institute of Science Bengaluru 560012, India}

\date{\today}

\begin{abstract}
The proposed spin-valve architecture comprises a two-dimensional graphene bilayer encapsulated between ferromagnetic insulating layers of ${\text{CrBr}}_3$. In this configuration, interfacial proximity exchange coupling plays a decisive role in determining the relative magnetization orientations of the graphene layers and modulating the spin-dependent transport characteristics of the heterostructure. Density functional theory (DFT) calculations were employed to investigate the spin-resolved electronic structure and interlayer coupling effects. The results reveal an induced band gap of approximately 56.87 meV at specific high-symmetry \textbf{k}-points, which becomes particularly prominent in the antiparallel magnetic configuration. This magnetization-dependent band modulation highlights the potential of the ${\text{CrBr}}_3$/graphene-bilayer/${\text{CrBr}}_3$ heterostructure as an efficient spin-filtering element in next-generation spintronic devices.
\end{abstract}

\maketitle

{\em Introduction--}Spin valves are crucial in spintronics, composed of layered ferromagnetic materials as conducting electrodes \cite{RevModPhys.81.109,PhysRevB.43.1297}. They are integral to modern electronics, leveraging giant magnetoresistance (MR) to switch between high and low resistance states \cite{monsma1995perpendicular,park2011aspinvalve,grunberg2008from,acs.nanolett.6b03052}. Understanding two-dimensional (2D) van der Waals (vdWs) systems has laid a robust foundation, offering finely tunable electronic properties in heterostructures~\cite{adma.202209113,acs.jpclett.3c03000,sciadv.abn1401,Flokstra2023,PhysRevLett.126.056803,PhysRevLett.128.167701} with materials like h-BN, TMDCs (e.g., MoS$_2$), graphene, and phosphorene \cite{Geim2013,Novoselovaac9439,PhysRevLett.114.066803,nat2019years,han2014graphene}. In vdW spin valves, the tunable spin proximity effect (SPE) can induce a metal-insulator transition in graphene under strong magnetic fields, potentially leading to Quantum Hall ferromagnetism \cite{PhysRevLett.95.146802,PhysRevB.82.161414,PhysRevLett.108.056802,PhysRevB.83.155447,PhysRevB.85.115439,acs.nanolett.3c02489,acs.nanolett.2c03113}. Advances with monolayers of ferromagnetic insulator CrBr$_3$ in vdWs devices have opened avenues for exploring spin-dependent transport phenomena \cite{Tombros2007,Song1214}.

Recently, a van der Waals (vdW) spin valve was demonstrated using bilayer graphene (BLG) between CrI$_3$ ferromagnetic insulators \cite{PhysRevLett.121.067701}. In our study, we aim to explore a similar configuration but replacing CrI$_3$ with CrBr$_3$, which has a slightly lower Curie temperature (T$_C$) of 33K \cite{PhysRevB.56.719} compared to CrI$_3$ (T$_C$ of 45K) \cite{Huang2017,PhysRevLett.121.067701}. CrBr$_3$ is notable for its heavy Cr atoms, suggesting potential for strong spin-orbit coupling (SOC) effects in graphene \cite{PhysRevB.84.195444,PhysRevLett.109.055502,acsami.4c01034}. Its hexagonal crystal structure aligns well with graphene's lattice, with a modest lattice mismatch of approximately 4.5 $\%$. This structural compatibility makes CrBr$_3$ promising for vdW heterostructures. The layered architecture of CrBr$_3$, bound by vdW forces, allows easy exfoliation into 2D monolayers while maintaining ferromagnetic insulating properties \cite{PhysRevLett.110.046603,Hallal_2017,acs.nanolett.9b00553}, enhancing the versatility of vdW spin valves. In CrBr$_3$ bilayers, weak interlayer coupling is conducive to controlling layer magnetizations independently, crucial for spintronic applications \cite{PhysRevB.84.195444,PhysRevLett.109.055502}. This exploration opens new avenues for studying spin-dependent phenomena in 2D vdW heterostructures.

Our study focuses on bilayer graphene, diverging from previous lateral spin valves that utilized monolayer graphene \cite{Tombros2007,han2014graphene}. Monolayer graphene has limited contact area with ferromagnetic electrodes, unlike spin filter tunnel junctions \cite{Klein1218,Song1214,Wang2018}. In contrast, our proposed van der Waals (vdW) spin valve integrates bilayer graphene between insulating magnetic layers \cite{PhysRevLett.121.067701,PhysRevLett.61.2472,PhysRevB.39.4828}. This design offers advantages such as increased contact area between the graphene and magnetic materials, potentially enhancing spin transport properties. Moreover, bilayer graphene's unique electronic structure and tunable properties enable precise control over spin-related phenomena. These attributes highlight the novelty and potential effectiveness of our vdW spin valve architecture.

\begin{figure*}
\centering
\includegraphics[width=14.0cm]{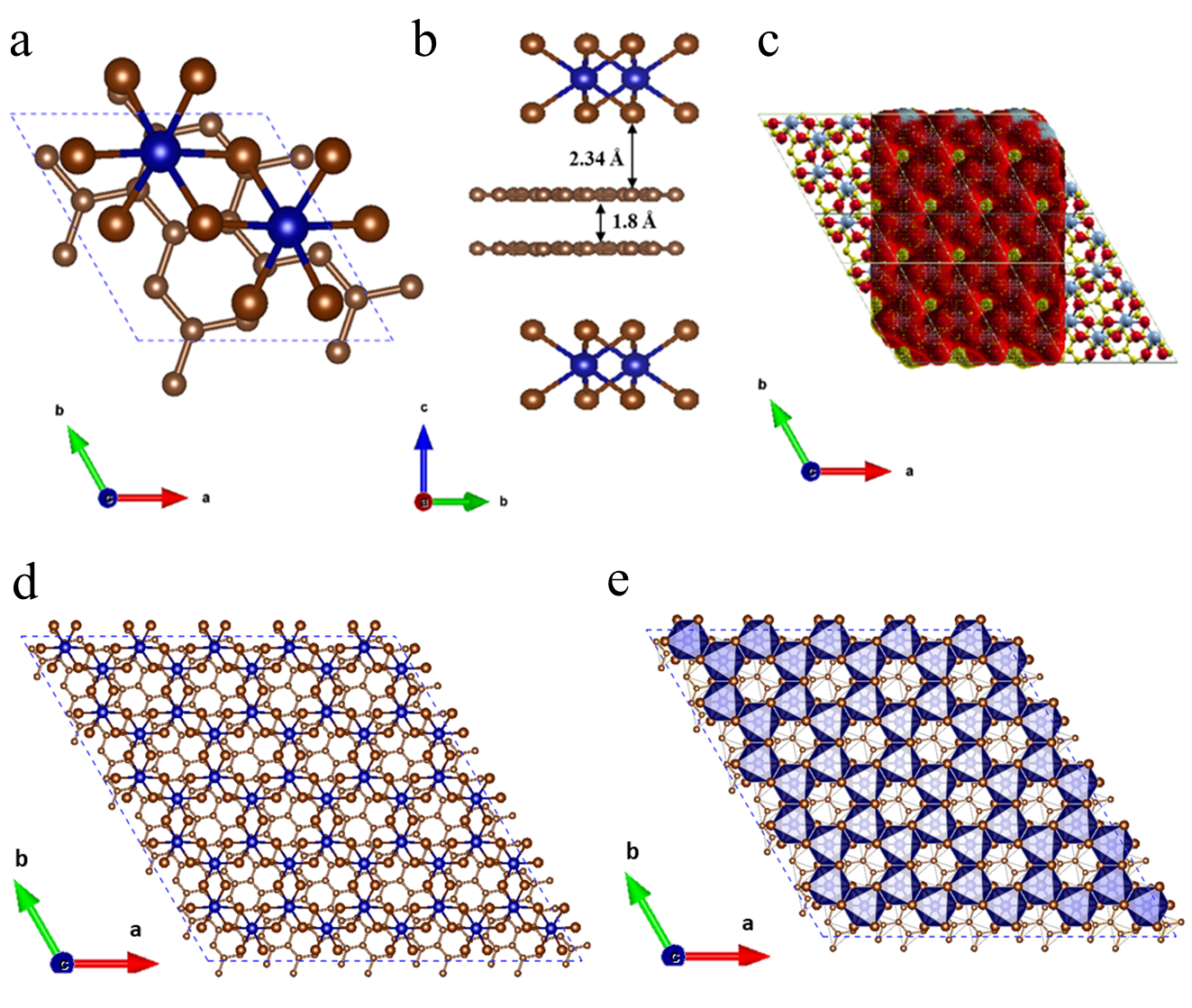}
  \caption{(a) Atomic structure of a CrBr$_3$-BLG-CrBr$_3$ heterostructure unit cell. Inter-planner spacing notation of the structure in (b) and molecular surface formation of the heterostructure showing distribution (c). Atomic structure of a CrBr$_3$-graphene bilayer-CrBr$_3$ heterostructure supercell of 5*5*1 (d). Polyhedra presentation of the same heterostructure showing the electron bond and distribution (e).}
  \label{sv-figure-1}
\end{figure*}

{\em Computational Details--}The geometry optimization and electronic structure calculations of the vdW heterostructure were conducted using the projector augmented wave (PAW) formalism \cite{PhysRevB.50.17953} within density functional theory (DFT) \cite{PhysRevB.54.11169}. The Perdew-Burke-Ernzerhof (PBE) generalized-gradient approximation (GGA) was employed to describe the exchange and correlation functional \cite{PhysRevLett.77.3865}. Specifically for the Cr 3d electrons, the GGA+U method was applied, with the effective \textit{on-site} Coulomb interaction U set to 3.0 eV and exchange interaction J to 0.9 eV \cite{PhysRevB.52.R5467}. For the monolayer CrBr$_3$, known for its accuracy but higher computational expense, the HSE06 hybrid functional \cite{10.1063/1.1564060} was adopted to achieve optimized geometry structures and electronic states. We design this heterostructure with a bilayer graphene (BLG) sheet surrounded by two monolayers of CrBr$_3$. For the bilayer structure, we used a unit cell composed of 4 layers: CrBr$_3$/BLG/CrBr$_3$. In the superlattice arrangement, we adjusted the in-plane lattice parameter of graphene while varying the $c$ lattice parameter to optimize the interlayer distance (2.34~\AA ~between CrBr$_3$/BLG and 1.8 \AA ~between BLG layers) and minimize the total energy. Each unit cell in the superlattice setup consisted of a 3$\times$3$\times$1 supercell, containing 14 carbon atoms per graphene layer, along with 2 chromium and 6 bromine atoms per CrBr$_3$ layer. Consequently, the sandwiched graphene bilayer structure encompassed a total of 44 atoms. Our calculations revealed that the spin of Cr atoms is $S=3/2$, residing within the $t{2g}$ bands. A plane-wave cutoff energy of 550 eV and a vacuum space larger than 22 {\AA} were implemented to prevent interactions between adjacent slabs. Grids of 9$\times$9$\times$1 and 27$\times$27$\times$1 Monkhorst-Pack k-point meshes were used for the vdW heterostructures and pure monolayer CrBr$_3$, respectively, to integrate over the first Brillouin zone. Here, we consider both the parallel configurations showing ferromagnetic (FM) and antiferromagentic (AFM) orientations. The vdW correction was incorporated using the Grimme (DFT-D2) method \cite{10.1002.20495} in the heterostructure calculations. Electronic band structures were also computed using the Quantum ESPRESSO suite \cite{Giannozzi_2009} to support and validate the computational findings.

\begin{figure}
\centering
\includegraphics[width=8.0cm]{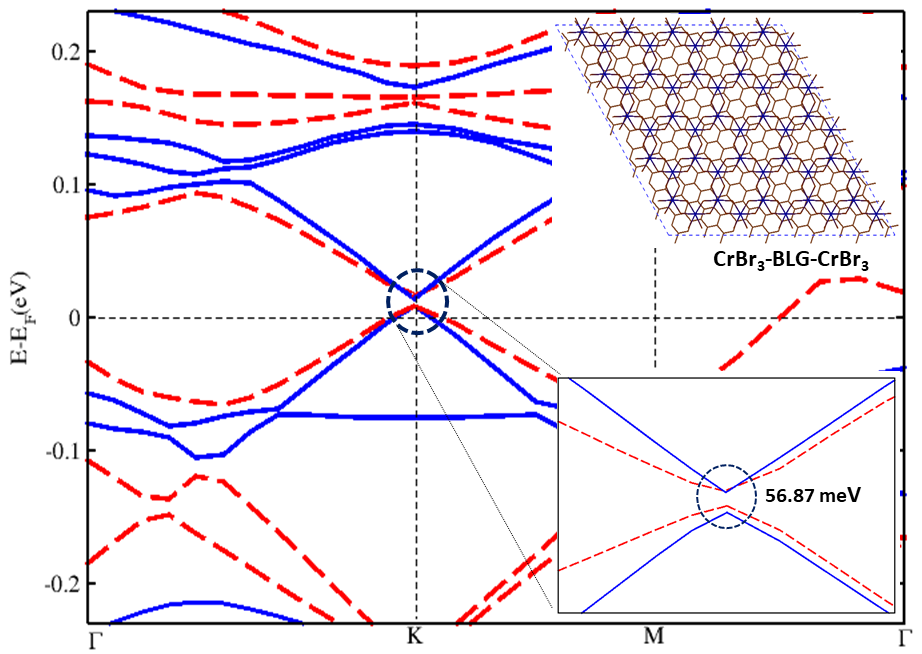}
  \caption{DFT band structure of the heterostructure with inset showing finite gap opening of 56.87 meV. Blue colour and red colour dotted lines present both majority and minority spins of both systems, respectively along highly symmetric \textit{k}-points.}
  \label{sv-figure-2}
\end{figure}

\begin{figure*}
\centering
\includegraphics[width=16.0cm,height=8.0cm]{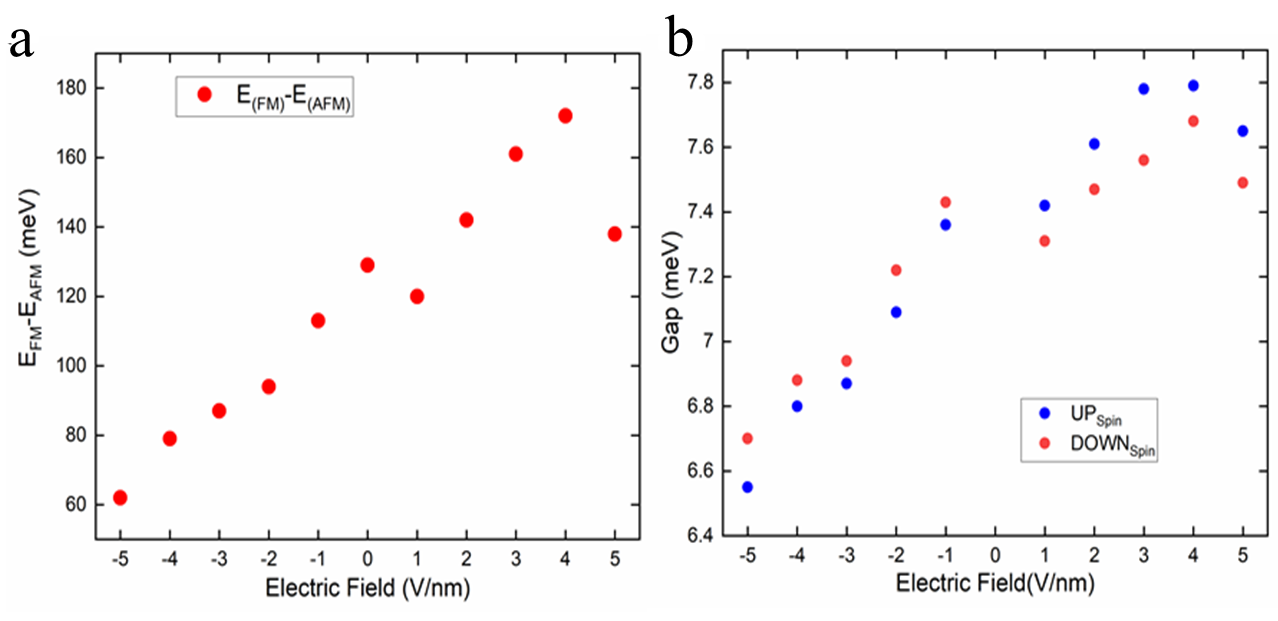}
 \caption{DFT calculation results in bilayer graphene sandwiched by CrBr$_3$. (a) The energy difference between ferromagnetic and antiferromagnetic states. (b) Spin-dependent gap as a function of U in presence of field.}
  \label{sv-figure-3}
\end{figure*}

{\em Results and Discussions--}The atomic configuration is depicted in Fig.\ref{sv-figure-1}(a-c), and the supercell architecture is illustrated in Fig.\ref{sv-figure-1}(d-e). The interlayer distance between the graphene sheet and CrBr$_3$ is 2.34 \AA, while the interplanar distance between graphene sheets is fixed at 1.80 \AA, characterized by van der Waals-type interlayer interactions. The total optimized energy is simulated as a function of interplanar distance to determine the optimal spacing for effective proximity coupling between the graphene sheet and CrBr$_3$, incorporating dispersion corrections \cite{monsma1995perpendicular}.

Dispersion correction has been integrated to accurately assess the bonding compatibility between the CrBr$_3$ layer and the graphene sheet, ensuring precise determination of the total energy relative to the interplanar spacing. In constructing graphene-on-metal systems as heterostructures, it is crucial to comprehend and minimize junction resistance originating from the metal-graphene interface during geometry optimization, alongside implementing dispersion corrections for van der Waals interacting systems \cite{park2011aspinvalve,grunberg2008from,han2014graphene}.

\begin{figure}
\centering
\includegraphics[width=9.0cm]{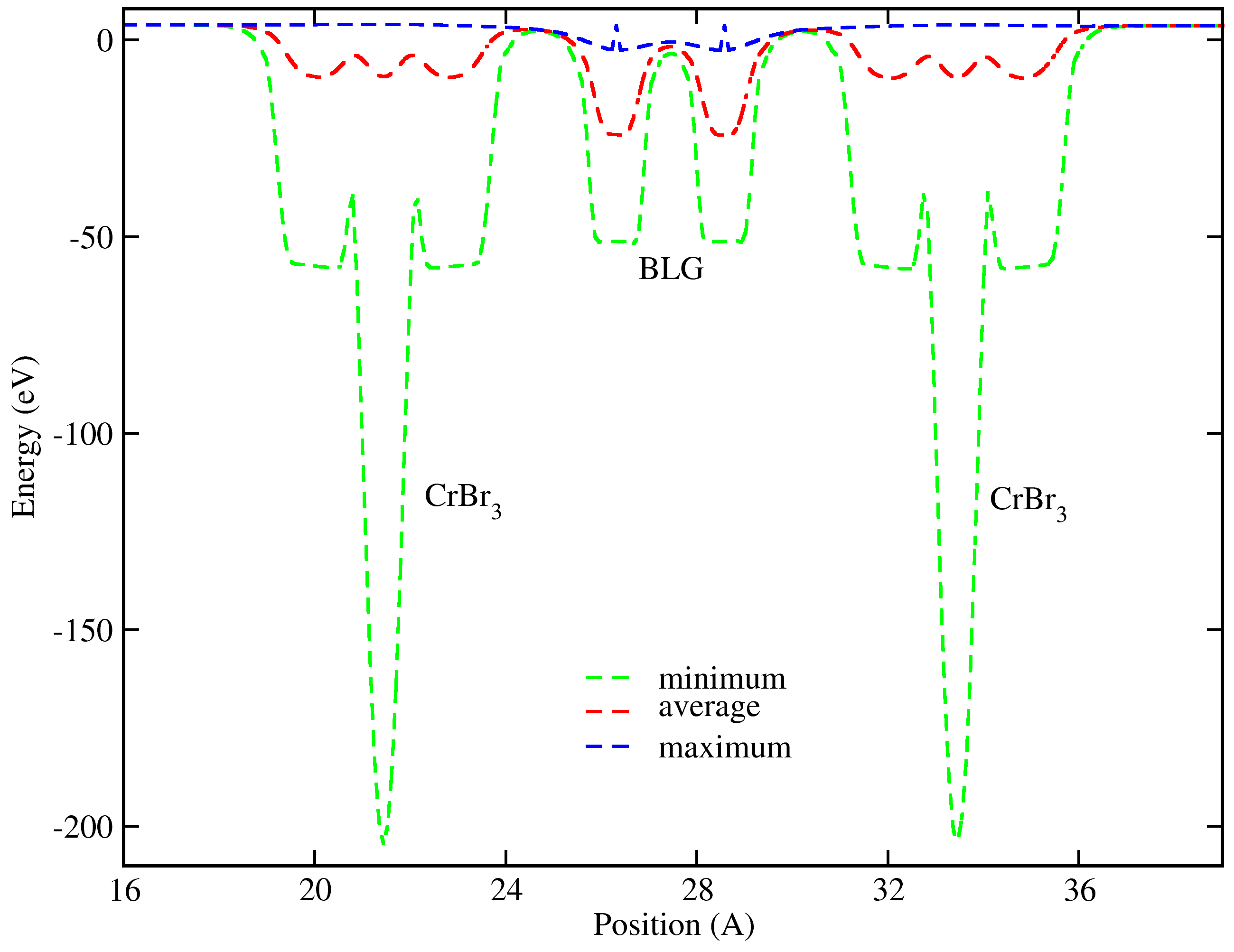}
  \caption{DFT local potential distribution of the CrBr$_3$-BLG-CrBr$_3$ heterostructure with contribution from both CrBr$_3$ monolayer and graphene bilayer along z-direction.}
  \label{sv-figure-4}
\end{figure}

The electronic band structure of the CrBr$_3$/BLG/CrBr$_3$ heterostructure at zero biasing is depicted in Fig.~\ref{sv-figure-2} under a ferromagnetic state. Analysis of the band structure reveals prominent graphene energy bands near the Fermi level, with an observed gap of 56.87 meV (inset of Fig.~\ref{sv-figure-2}. This characteristic illustrates well-organized energy levels at the \textit{K} point, accounting for spin degeneracy. Consequently, the proximity magnetism exerted by the CrBr$_3$ layers on the BLG system manifests a discernible AFM ground state (depicted in Fig.~\ref{sv-figure-2}, unlike FM state. To support the electronic band structure analysis, the band structure of our model vdW spin valve is depicted in Figure \ref{sv-figure-2} alongside the $e_g$ bands. In this scenario, the shape of the graphene bilayer bands remains unchanged in the minority spin channel, while significant hybridization introduces a gap in the majority spin channel just above the Fermi energy, positioned below the \textit{k} point. In contrast, both spin channels of the graphene bilayer undergo hybridization with the narrow $e_g$ bands of CrBr$_3$. This analysis suggests a notably higher \textit{in-plane} conductance in the AFM alignment based solely on this spintronics perspective. 

We assume that the magnetizations of both the top and bottom layers are aligned along the same axis during the calculations. The magnetization of the bottom layer is assumed to be fixed, leading to a constant spin-dependent potential of $\Sigma \delta$, where $\Sigma$ represents the spin projection along this axis. For the top layer, the spin-dependent potential is given by $\epsilon \Sigma \delta$, where $\epsilon = \pm 1$ indicates the relative orientation of the top layer's magnetization with respect to the bottom layer. Specifically, $\epsilon = +1$ corresponds to a parallel FM alignment, while $\epsilon = -1$ corresponds to an antiparallel AFM alignment. In this case, the Bloch Hamiltonian for spin $\Sigma$ states is expressed as follows,:
\begin{eqnarray}
{\cal H}_{\Sigma}(\vec{k})=\left(
\begin{array}{cccc}
\epsilon\Sigma \frac{\delta}{2} & f(\vec{k}) & 0& 0 \\
f^*(\vec{k}) & \epsilon\Sigma \frac{\delta}{2} & \Gamma  & 0 \\
0 & \Gamma  &  \Sigma \frac{\delta}{2} &  f(\vec{k})\\
0 & 0 &  f(\vec{k})^* & \Sigma \frac{\delta}{2}
\end{array}
\right)
\label{hmodel}
\end{eqnarray}
Here, $f(\vec{k}) = t\left(1 + e^{i\vec{k}\cdot\vec{a}_1} + e^{i\vec{k}\cdot\vec{a}_2}\right)$ and $\Gamma$ denote the intralayer and interlayer hopping matrix elements, respectively. The spin-resolved energy bands near the Dirac {\em K} point are illustrated in Fig.~\ref{sv-figure-2} for the spin valve architecture. In the FM alignment ($\epsilon = +1$), the BLG system exhibits spin-split bands and remains in a metallic state, meaning it has a finite density of states at the Fermi energy. Conversely, in the AFM alignment ($\epsilon = -1$), a band gap (56.87 meV) opens at the Dirac {\em K} point. Thus, the BLG spin valve can act as a conductor in the FM state or as a gapped insulator at the Dirac {\em K} point in the AFM state. 

These results highlight the potential of using a vdW magnetic system to explore quantum Hall effects, diverging from conventional ferromagnetic layer systems. Examining the band structure reveals fluctuations in the \textit{split-off} energy gap between the ferromagnetic and antiferromagnetic states (illustrated in Fig.\ref{sv-figure-3}(a)), along with the spin-dependent gap as a function of \textit{U} across bias voltages from 0 to ±5 V/nm (shown in Fig.\ref{sv-figure-3}(b)). This analysis demonstrates pronounced non-linearity under biasing conditions, indicative of interlayer charge polarization. Moreover, the mixed (both linear and nonlinear) behavior prompts further investigation into charge transmission spectra to understand the proximity effect in heterostructures and the role of interlayer polarization for potential device applications (as depicted in the inset heterostructure in Fig.~\ref{sv-figure-2}). We now explore the origin of gap formation in the graphene bilayer within an antiferromagnetic aligned vdW spin valve. In this setup, the Hamiltonian for each spin channel mirrors the model of a graphene bilayer under an off-plane electric field, which is known to induce a band gap in the structure \cite{Ohta951,PhysRevB.74.161403}. Notably, within the spin valve configuration, the effective electric field exhibits opposite signs for different spin orientations, $E_{\rm eff}\propto \sigma \Delta$. The spin projection of the antiferromagnetic bands in the top and bottom layers, as depicted in Fig. \ref{sv-figure-3}(b), distinctly manifests a spin dipole effect: for a given spin direction, there arises a charge imbalance induced by the exchange interaction with the magnetic layers, precisely offset by the opposite spin direction.

We explore the energy states around the Fermi energy ($E_F$) of a graphene bilayer, where the valence and conduction parabolic bands converge. This region shows a slight band splitting, indicative of parabolic electrons and holes in the BLG structure, with a split of 56.87 meV observed for the AFM case (as depicted in the inset of Fig. \ref{sv-figure-2}). Similarly, there is a distinct spin splitting of the bands near the \textit{k} point in this configuration, with the same magnitude as the AFM gap. This consistency supports the DFT findings for bilayer graphene sandwiched between two CrBr$_3$ layers, consistent with theoretical predictions. However, due to electron transfer from graphene to CrBr$_3$, the Fermi energy does not precisely align at the \textit{k} point. It's noteworthy that despite this charge transfer, CrBr$_3$ exhibits very limited in-plane conductance due to the minimal dispersion of occupied states, as evidenced by the local potential distribution plot (see Fig. \ref{sv-figure-4}), showing the contributions of BLG and CrBr$_3$ layers.

{\em Summary--}We have proposed a novel spin valve design where a non-magnetic 2D crystal is sandwiched between two ferromagnetic insulators. Its \textit{in-plane} conductance, mediated through nanoscale transport, is controlled by the spin proximity effect. We anticipate that our work will inspire experimentalists to explore van der Waals (vdW) in-plane spin valves further. This includes investigating other materials for the non-magnetic layers, such as superconductors, as well as alternative magnetic layers, like bulk ferromagnets, for which the proposed spin valve effect should also be applicable. \\

{\em keywords--} vdW heterostructure, ${CrBr}_3$, Spin proximity effect, Spintronics, Density functional theory

\bibliography{spin_valve}

@article{monsma1995perpendicular,
  title = {Perpendicular Hot Electron Spin-Valve Effect in a New Magnetic Field Sensor: The Spin-Valve Transistor},
  author = {Monsma, D. J. and Lodder, J. C. and Popma, Th. J. A. and Dieny, B.},
  journal = {Phys. Rev. Lett.},
  volume = {74},
  issue = {26},
  pages = {5260--5263},
  numpages = {0},
  year = {1995},
  month = {Jun},
  publisher = {American Physical Society},
}

@Article{park2011aspinvalve,
author={Park, B. G.
and Wunderlich, J.
and Mart{\'i}, X.
and Hol{\'y}, V.
and Kurosaki, Y.
and Yamada, M.
and Yamamoto, H.
and Nishide, A.
and Hayakawa, J.
and Takahashi, H.
and Shick, A. B.
and Jungwirth, T.},
title={A spin-valve-like magnetoresistance of an antiferromagnet-based tunnel junction},
journal={Nature Materials},
year={2011},
month={May},
day={01},
volume={10},
number={5},
pages={347-351},
}

@article{grunberg2008from,
  title = {Nobel Lecture: From spin waves to giant magnetoresistance and beyond},
  author = {Gr\"unberg, Peter A.},
  journal = {Rev. Mod. Phys.},
  volume = {80},
  issue = {4},
  pages = {1531--1540},
  numpages = {0},
  year = {2008},
  month = {Dec},
  publisher = {American Physical Society},
}

@Article{nat2019years,
title={15 years of graphene electronics},
journal={Nature Electronics},
year={2019},
month={Sep},
day={01},
volume={2},
number={9},
pages={369-369},
}

@Article{han2014graphene,
author={Han, Wei
and Kawakami, Roland K.
and Gmitra, Martin
and Fabian, Jaroslav},
title={Graphene spintronics},
journal={Nature Nanotechnology},
year={2014},
month={Oct},
day={01},
volume={9},
number={10},
pages={794-807},
}

@article{PhysRevLett.95.146802,
  title = {${Z}_{2}$ Topological Order and the Quantum Spin Hall Effect},
  author = {Kane, C. L. and Mele, E. J.},
  journal = {Phys. Rev. Lett.},
  volume = {95},
  issue = {14},
  pages = {146802},
  numpages = {4},
  year = {2005},
  month = {Sep},
  publisher = {American Physical Society},
  doi = {10.1103/PhysRevLett.95.146802},
  url = {https://link.aps.org/doi/10.1103/PhysRevLett.95.146802}
}

@article{PhysRevB.82.161414,
  title = {Quantum anomalous Hall effect in graphene from Rashba and exchange effects},
  author = {Qiao, Zhenhua and Yang, Shengyuan A. and Feng, Wanxiang and Tse, Wang-Kong and Ding, Jun and Yao, Yugui and Wang, Jian and Niu, Qian},
  journal = {Phys. Rev. B},
  volume = {82},
  issue = {16},
  pages = {161414},
  numpages = {4},
  year = {2010},
  month = {Oct},
  publisher = {American Physical Society},
  doi = {10.1103/PhysRevB.82.161414},
  url = {https://link.aps.org/doi/10.1103/PhysRevB.82.161414}
}

@article{PhysRevB.50.17953,
  title = {Projector augmented-wave method},
  author = {Bl\"ochl, P. E.},
  journal = {Phys. Rev. B},
  volume = {50},
  issue = {24},
  pages = {17953--17979},
  numpages = {0},
  year = {1994},
  month = {Dec},
  publisher = {American Physical Society},
  doi = {10.1103/PhysRevB.50.17953},
  url = {https://link.aps.org/doi/10.1103/PhysRevB.50.17953}
}

@article{PhysRevB.54.11169,
  title = {Efficient iterative schemes for ab initio total-energy calculations using a plane-wave basis set},
  author = {Kresse, G. and Furthm\"uller, J.},
  journal = {Phys. Rev. B},
  volume = {54},
  issue = {16},
  pages = {11169--11186},
  numpages = {0},
  year = {1996},
  month = {Oct},
  publisher = {American Physical Society},
  doi = {10.1103/PhysRevB.54.11169},
  url = {https://link.aps.org/doi/10.1103/PhysRevB.54.11169}
}

@article{PhysRevLett.77.3865,
  title = {Generalized Gradient Approximation Made Simple},
  author = {Perdew, John P. and Burke, Kieron and Ernzerhof, Matthias},
  journal = {Phys. Rev. Lett.},
  volume = {77},
  issue = {18},
  pages = {3865--3868},
  numpages = {0},
  year = {1996},
  month = {Oct},
  publisher = {American Physical Society},
  doi = {10.1103/PhysRevLett.77.3865},
  url = {https://link.aps.org/doi/10.1103/PhysRevLett.77.3865}
}

@article{PhysRevB.52.R5467,
  title = {Density-functional theory and strong interactions: Orbital ordering in Mott-Hubbard insulators},
  author = {Liechtenstein, A. I. and Anisimov, V. I. and Zaanen, J.},
  journal = {Phys. Rev. B},
  volume = {52},
  issue = {8},
  pages = {R5467--R5470},
  numpages = {0},
  year = {1995},
  month = {Aug},
  publisher = {American Physical Society},
  doi = {10.1103/PhysRevB.52.R5467},
  url = {https://link.aps.org/doi/10.1103/PhysRevB.52.R5467}
}

@article{10.1063/1.1564060,
author = {Heyd,Jochen  and Scuseria,Gustavo E.  and Ernzerhof,Matthias },
title = {Hybrid functionals based on a screened Coulomb potential},
journal = {The Journal of Chemical Physics},
volume = {118},
number = {18},
pages = {8207-8215},
year = {2003},
doi = {10.1063/1.1564060},
URL = {https://doi.org/10.1063/1.1564060},
}

@article{10.1002.20495,
author = {Grimme, Stefan},
title = {Semiempirical GGA-type density functional constructed with a long-range dispersion correction},
journal = {Journal of Computational Chemistry},
volume = {27},
number = {15},
pages = {1787-1799},
doi = {https://doi.org/10.1002/jcc.20495},
url = {https://onlinelibrary.wiley.com/doi/abs/10.1002/jcc.20495},
year = {2006}
}

@article{PhysRevLett.108.056802,
  title = {Electrically Tunable Quantum Anomalous Hall Effect in Graphene Decorated by $5d$ Transition-Metal Adatoms},
  author = {Zhang, Hongbin and Lazo, Cesar and Bl\"ugel, Stefan and Heinze, Stefan and Mokrousov, Yuriy},
  journal = {Phys. Rev. Lett.},
  volume = {108},
  issue = {5},
  pages = {056802},
  numpages = {5},
  year = {2012},
  month = {Feb},
  publisher = {American Physical Society},
  doi = {10.1103/PhysRevLett.108.056802},
  url = {https://link.aps.org/doi/10.1103/PhysRevLett.108.056802}
}

@article{PhysRevB.83.155447,
  title = {Quantum anomalous Hall effect in single-layer and bilayer graphene},
  author = {Tse, Wang-Kong and Qiao, Zhenhua and Yao, Yugui and MacDonald, A. H. and Niu, Qian},
  journal = {Phys. Rev. B},
  volume = {83},
  issue = {15},
  pages = {155447},
  numpages = {8},
  year = {2011},
  month = {Apr},
  publisher = {American Physical Society},
  doi = {10.1103/PhysRevB.83.155447},
  url = {https://link.aps.org/doi/10.1103/PhysRevB.83.155447}
}

@article{PhysRevLett.109.055502,
  title = {Valley-Polarized Metals and Quantum Anomalous Hall Effect in Silicene},
  author = {Ezawa, Motohiko},
  journal = {Phys. Rev. Lett.},
  volume = {109},
  issue = {5},
  pages = {055502},
  numpages = {5},
  year = {2012},
  month = {Aug},
  publisher = {American Physical Society},
  doi = {10.1103/PhysRevLett.109.055502},
  url = {https://link.aps.org/doi/10.1103/PhysRevLett.109.055502}
}

@article{PhysRevB.84.195444,
  title = {Engineering quantum anomalous/valley Hall states in graphene via metal-atom adsorption: An ab-initio study},
  author = {Ding, Jun and Qiao, Zhenhua and Feng, Wanxiang and Yao, Yugui and Niu, Qian},
  journal = {Phys. Rev. B},
  volume = {84},
  issue = {19},
  pages = {195444},
  numpages = {9},
  year = {2011},
  month = {Nov},
  publisher = {American Physical Society},
  doi = {10.1103/PhysRevB.84.195444},
  url = {https://link.aps.org/doi/10.1103/PhysRevB.84.195444}
}

@article{PhysRevB.85.115439,
  title = {Microscopic theory of quantum anomalous Hall effect in graphene},
  author = {Qiao, Zhenhua and Jiang, Hua and Li, Xiao and Yao, Yugui and Niu, Qian},
  journal = {Phys. Rev. B},
  volume = {85},
  issue = {11},
  pages = {115439},
  numpages = {10},
  year = {2012},
  month = {Mar},
  publisher = {American Physical Society},
  doi = {10.1103/PhysRevB.85.115439},
  url = {https://link.aps.org/doi/10.1103/PhysRevB.85.115439}
}

@Article{Geim2013,
author={Geim, A. K.
and Grigorieva, I. V.},
title={Van der Waals heterostructures},
journal={Nature},
year={2013},
month={Jul},
day={01},
volume={499},
number={7459},
pages={419-425},
issn={1476-4687},
doi={10.1038/nature12385},
url={https://doi.org/10.1038/nature12385}
}

@article{PhysRevB.56.719,
  title = {rf susceptibility of single-crystal ${\mathrm{CrBr}}_{3}$ near the Curie temperature},
  author = {Alyoshin, V. A. and Berezin, V. A. and Tulin, V. A.},
  journal = {Phys. Rev. B},
  volume = {56},
  issue = {2},
  pages = {719--725},
  numpages = {0},
  year = {1997},
  month = {Jul},
  publisher = {American Physical Society},
  doi = {10.1103/PhysRevB.56.719},
  url = {https://link.aps.org/doi/10.1103/PhysRevB.56.719}
}

@article{acs.nanolett.9b00553,
author = {Zhang, Zhaowei and Shang, Jingzhi and Jiang, Chongyun and Rasmita, Abdullah and Gao, Weibo and Yu, Ting},
title = {Direct Photoluminescence Probing of Ferromagnetism in Monolayer Two-Dimensional CrBr3},
journal = {Nano Letters},
volume = {19},
number = {5},
pages = {3138-3142},
year = {2019},
doi = {10.1021/acs.nanolett.9b00553},
URL = {https://doi.org/10.1021/acs.nanolett.9b00553},
}

@article{RevModPhys.81.109,
  title = {The electronic properties of graphene},
  author = {Castro Neto, A. H. and Guinea, F. and Peres, N. M. R. and Novoselov, K. S. and Geim, A. K.},
  journal = {Rev. Mod. Phys.},
  volume = {81},
  issue = {1},
  pages = {109--162},
  numpages = {0},
  year = {2009},
  month = {Jan},
  publisher = {American Physical Society},
  doi = {10.1103/RevModPhys.81.109},
  url = {https://link.aps.org/doi/10.1103/RevModPhys.81.109}
}

@article{PhysRevB.43.1297,
  title = {Giant magnetoresistive in soft ferromagnetic multilayers},
  author = {Dieny, B. and Speriosu, V. S. and Parkin, S. S. P. and Gurney, B. A. and Wilhoit, D. R. and Mauri, D.},
  journal = {Phys. Rev. B},
  volume = {43},
  issue = {1},
  pages = {1297--1300},
  numpages = {0},
  year = {1991},
  month = {Jan},
  publisher = {American Physical Society},
  doi = {10.1103/PhysRevB.43.1297},
  url = {https://link.aps.org/doi/10.1103/PhysRevB.43.1297}
}

@article {Novoselovaac9439,
	author = {Novoselov, K. S. and Mishchenko, A. and Carvalho, A. and Castro Neto, A. H.},
	title = {2D materials and van der Waals heterostructures},
	volume = {353},
	number = {6298},
	elocation-id = {aac9439},
	year = {2016},
	doi = {10.1126/science.aac9439},
	publisher = {American Association for the Advancement of Science},
	URL = {https://science.sciencemag.org/content/353/6298/aac9439},
	journal = {Science}
}

@article{PhysRevLett.114.066803,
  title = {van der Waals Heterostructure of Phosphorene and Graphene: Tuning the Schottky Barrier and Doping by Electrostatic Gating},
  author = {Padilha, J. E. and Fazzio, A. and da Silva, Ant\^onio J. R.},
  journal = {Phys. Rev. Lett.},
  volume = {114},
  issue = {6},
  pages = {066803},
  numpages = {5},
  year = {2015},
  month = {Feb},
  publisher = {American Physical Society},
  doi = {10.1103/PhysRevLett.114.066803},
  url = {https://link.aps.org/doi/10.1103/PhysRevLett.114.066803}
}

@article{acs.nanolett.6b03052,
author = {Lee, Jae-Ung and Lee, Sungmin and Ryoo, Ji Hoon and Kang, Soonmin and Kim, Tae Yun and Kim, Pilkwang and Park, Cheol-Hwan and Park, Je-Geun and Cheong, Hyeonsik},
title = {Ising-Type Magnetic Ordering in Atomically Thin FePS3},
journal = {Nano Letters},
volume = {16},
number = {12},
pages = {7433-7438},
year = {2016},
doi = {10.1021/acs.nanolett.6b03052},
note ={PMID: 27960508},
URL = {https://doi.org/10.1021/acs.nanolett.6b03052},
eprint = {https://doi.org/10.1021/acs.nanolett.6b03052}
}

@Article{Huang2017,
author={Huang, Bevin
and Clark, Genevieve
and Navarro-Moratalla, Efr{\'e}n
and Klein, Dahlia R.
and Cheng, Ran
and Seyler, Kyle L.
and Zhong, Ding
and Schmidgall, Emma
and McGuire, Michael A.
and Cobden, David H.
and Yao, Wang
and Xiao, Di
and Jarillo-Herrero, Pablo
and Xu, Xiaodong},
title={Layer-dependent ferromagnetism in a van der Waals crystal down to the monolayer limit},
journal={Nature},
year={2017},
month={Jun},
day={01},
volume={546},
number={7657},
pages={270-273},
issn={1476-4687},
doi={10.1038/nature22391},
url={https://doi.org/10.1038/nature22391}
}

@article {Song1214,
	author = {Song, Tiancheng and Cai, Xinghan and Tu, Matisse Wei-Yuan and Zhang, Xiaoou and Huang, Bevin and Wilson, Nathan P. and Seyler, Kyle L. and Zhu, Lin and Taniguchi, Takashi and Watanabe, Kenji and McGuire, Michael A. and Cobden, David H. and Xiao, Di and Yao, Wang and Xu, Xiaodong},
	title = {Giant tunneling magnetoresistance in spin-filter van der Waals heterostructures},
	volume = {360},
	number = {6394},
	pages = {1214--1218},
	year = {2018},
	doi = {10.1126/science.aar4851},
	publisher = {American Association for the Advancement of Science},
	URL = {https://science.sciencemag.org/content/360/6394/1214},
	journal = {Science}
}

@article {Klein1218,
	author = {Klein, D. R. and MacNeill, D. and Lado, J. L. and Soriano, D. and Navarro-Moratalla, E. and Watanabe, K. and Taniguchi, T. and Manni, S. and Canfield, P. and Fern{\'a}ndez-Rossier, J. and Jarillo-Herrero, P.},
	title = {Probing magnetism in 2D van der Waals crystalline insulators via electron tunneling},
	volume = {360},
	number = {6394},
	pages = {1218--1222},
	year = {2018},
	doi = {10.1126/science.aar3617},
	publisher = {American Association for the Advancement of Science},
	URL = {https://science.sciencemag.org/content/360/6394/1218},
	journal = {Science}
}

@Article{Tombros2007,
author={Tombros, Nikolaos
and Jozsa, Csaba
and Popinciuc, Mihaita
and Jonkman, Harry T.
and van Wees, Bart J.},
title={Electronic spin transport and spin precession in single graphene layers at room temperature},
journal={Nature},
year={2007},
month={Aug},
day={01},
volume={448},
number={7153},
pages={571-574},
issn={1476-4687},
doi={10.1038/nature06037},
url={https://doi.org/10.1038/nature06037}
}

@Article{Wang2018,
author={Wang, Zhe
and Guti{\'e}rrez-Lezama, Ignacio
and Ubrig, Nicolas
and Kroner, Martin
and Gibertini, Marco
and Taniguchi, Takashi
and Watanabe, Kenji
and Imamo{\u{g}}lu, Ata{\c{c}}
and Giannini, Enrico
and Morpurgo, Alberto F.},
title={Very large tunneling magnetoresistance in layered magnetic semiconductor CrI3},
journal={Nature Communications},
year={2018},
month={Jun},
day={28},
volume={9},
number={1},
pages={2516},
doi={10.1038/s41467-018-04953-8},
url={https://doi.org/10.1038/s41467-018-04953-8}
}

@article{PhysRevLett.61.2472,
  title = {Giant Magnetoresistance of (001)Fe/(001)Cr Magnetic Superlattices},
  author = {Baibich, M. N. and Broto, J. M. and Fert, A. and Van Dau, F. Nguyen and Petroff, F. and Etienne, P. and Creuzet, G. and Friederich, A. and Chazelas, J.},
  journal = {Phys. Rev. Lett.},
  volume = {61},
  issue = {21},
  pages = {2472--2475},
  numpages = {0},
  year = {1988},
  month = {Nov},
  publisher = {American Physical Society},
  doi = {10.1103/PhysRevLett.61.2472},
  url = {https://link.aps.org/doi/10.1103/PhysRevLett.61.2472}
}

@article{PhysRevB.39.4828,
  title = {Enhanced magnetoresistance in layered magnetic structures with antiferromagnetic interlayer exchange},
  author = {Binasch, G. and Gr\"unberg, P. and Saurenbach, F. and Zinn, W.},
  journal = {Phys. Rev. B},
  volume = {39},
  issue = {7},
  pages = {4828--4830},
  numpages = {0},
  year = {1989},
  month = {Mar},
  publisher = {American Physical Society},
  doi = {10.1103/PhysRevB.39.4828},
  url = {https://link.aps.org/doi/10.1103/PhysRevB.39.4828}
}

@article{PhysRevLett.110.046603,
  title = {Proximity Effects Induced in Graphene by Magnetic Insulators: First-Principles Calculations on Spin Filtering and Exchange-Splitting Gaps},
  author = {Yang, H. X. and Hallal, A. and Terrade, D. and Waintal, X. and Roche, S. and Chshiev, M.},
  journal = {Phys. Rev. Lett.},
  volume = {110},
  issue = {4},
  pages = {046603},
  numpages = {5},
  year = {2013},
  month = {Jan},
  publisher = {American Physical Society},
  doi = {10.1103/PhysRevLett.110.046603},
  url = {https://link.aps.org/doi/10.1103/PhysRevLett.110.046603}
}

@article{Hallal_2017,
	doi = {10.1088/2053-1583/aa6663},
	url = {https://doi.org/10.1088/2053-1583/aa6663},
	year = 2017,
	month = {apr},
	publisher = {{IOP} Publishing},
	volume = {4},
	number = {2},
	pages = {025074},
	author = {Ali Hallal and Fatima Ibrahim and Hongxin Yang and Stephan Roche and Mairbek Chshiev},
	title = {Tailoring magnetic insulator proximity effects in graphene: first-principles calculations},
	journal = {2D Materials},

}

@article{PhysRevLett.121.067701,
  title = {Van der Waals Spin Valves},
  author = {Cardoso, C. and Soriano, D. and Garc\'{\i}a-Mart\'{\i}nez, N. A. and Fern\'andez-Rossier, J.},
  journal = {Phys. Rev. Lett.},
  volume = {121},
  issue = {6},
  pages = {067701},
  numpages = {5},
  year = {2018},
  month = {Aug},
  publisher = {American Physical Society},
  doi = {10.1103/PhysRevLett.121.067701},
  url = {https://link.aps.org/doi/10.1103/PhysRevLett.121.067701}
}

@article {Ohta951,
	author = {Ohta, Taisuke and Bostwick, Aaron and Seyller, Thomas and Horn, Karsten and Rotenberg, Eli},
	title = {Controlling the Electronic Structure of Bilayer Graphene},
	volume = {313},
	number = {5789},
	pages = {951--954},
	year = {2006},
	doi = {10.1126/science.1130681},
	publisher = {American Association for the Advancement of Science},
	URL = {https://science.sciencemag.org/content/313/5789/951},
	journal = {Science}
}

\end{document}